\documentclass[12pt,a4paper]{article}
\usepackage{jheppub_kim}
 \usepackage{slashed}
\usepackage{longtable}
\usepackage{epsfig}
\usepackage{amssymb,amsmath}
\usepackage{amsthm}
\usepackage{graphicx}
 \usepackage{graphics}
\usepackage[hang,nooneline,scriptsize]{subfigure}
\usepackage{subfigure}
\usepackage{array}
\begin{document}
  
\title{Shadow of the $5D$  Reissner-Nordstr\"om AdS Black Hole}
 
\author[a] {Surajit Mandal}
 \author [b,c,]{ Sudhaker Upadhyay\footnote{Visiting Associate,  IUCAA  Pune, Maharashtra-411007}}
    \author[d,e]{Yerlan Myrzakulov}
    \author[d,e]{Gulmira Yergaliyeva} 
		\affiliation[a]{Department of Physics, Jadavpur University, Kolkata, West Bengal 700032, India }
	 \affiliation[b]{Department of Physics, K. L. S. College, Magadh University, Nawada, Bihar 805110, India}
 
  \affiliation[c]{School of Physics, Damghan University, Damghan, 3671641167, Iran}

\affiliation[d]{Department of General \& Theoretical Physics, LN Gumilyov Eurasian National University, Nur-Sultan, 010008, Kazakhstan}
\affiliation[e]{ Ratbay Myrzakulov Eurasian International Center for Theoretical Physics, Nur-Sultan, 010009, Kazakhstan}

\emailAdd{surajitmandalju@gmail.com}	
 \emailAdd{sudhakerupadhyay@gmail.com} \emailAdd{sudhaker@associates.iucaa.in}  
 \emailAdd{ymyrzakulov@gmail.com} 
  \emailAdd{gyergaliyeva1171@gmail.com} 
\abstract{We discuss the shadow cast by the  charged Reissner-Nordstr\"om (RN) AdS black hole. With the help of Killing equation and Hamilton-Jacobi
equation, we calculate  the geodesic equations for null particle. With the help of  geodesics of null particle, we then determine the celestial coordinates ($\alpha$, $\beta$) and the  shadow radius of the RN AdS black hole. We present a graphical analysis of  the black hole shadow  and find that shadow is a perfectly dark   circle. The impacts of  charge and cosmological constant of the RN AdS black hole on the radius  of  shadow  are also presented. In this connection, radius of the  shadow is a decreasing function of the charge. Furthermore, we   study  the effects of plasma medium   on the RN AdS  black hole shadow. Here, we  find that radius of circular shadow   increases with increasing plasma parameter. In addition, we also discuss the energy emission rate of RN AdS  black hole.  The  effects of parameters like charge, cosmological constant and plasma parameter on energy emission rate are analyzed graphically.}	
	\maketitle

\section{Introduction}
The study of black hole is an interesting topic of physics since the discovery of Einstein's theory of general relativity. Nowadays, it is sharply presumed that there exist black holes at the center   of most of the galaxies, for example, it is mostly accepted that supermassive black hole Sagittarius $A^{\star}$ or Sgr $A^{\star}$ is present in the galactic
center of Milky Way \cite{si1, si2} and recently the shadow of this supermassive black hole has been noticed by event horizon telescope (EHT) group \cite{si3}. Any objects moving near the black hole  has a intense gravitational attraction and,  within a critical radius, must falls into the black hole. This is happen due to strong gravitational lensing. 

If an illuminated
source of photon is present just behind the black hole, then photon particles can move in the vicinity of black holes and can make a shadow at a plane which might be visible by a distant viewer. The shadow  of the black hole was first studied by Bardeen   \cite{si4}. It has been found  that the spherically symmetric static (Schwarzschild) black holes can cast a shadow of circular shape \cite{si5, si6}, whereas the  rotating (spinning) black holes cast   deformed shadows from circular shape \cite{si7, si8, si8a, si8b, si8c, si8d}. Nowadays, the study of shadow cast of black hole has becomes an active area of research   and received a momentous attention \cite{si9}. The shadow of Schwarzschild black hole has been studied in \cite{si10, si11}. The   rotating black hole shadow along with gravitomagnetic and electric charge can be found in \cite{si12}. The gravitational lensing together with optical phenomena for the Janis-Newman-Winicour
  \cite{si13}, the Kerr-Newman and   the Kerr-Newman-AdS spacetimes have been studied in Ref. \cite{si14}. Peoples have investigated the shadow of other black holes, such as regular black holes \cite{si15, si16, si17}, Kerr black holes \cite{si4}, Kerr-Newman black holes \cite{si18} and  multi-black holes \cite{si19}. Shadow of black holes in a modified gravity due to extended Chern-Simons and  Einstein-Maxwell dilaton gravity can be found in \cite{si20}. However, the shadow in Randall-Sundrum braneworld case is studied in \cite{si21}. Recently,  \"Ovg\"un et al.   studied the black hole shadow of noncommutative black holes in Rastall gravity \cite{s11}. Testing of generalized Einstein-Cartan-Kibble-Sciama gravity using shadow cast was made in \cite{si22}.   Jusufi et al. \cite{si22a} studied the shadows of $5D$   charged Bardeen black holes. Shadow cast of rotating braneworld black holes
with a cosmological constant was made in \cite{si22b}.
 
The black holes solution in higher than four dimensions are of momentous  importance.  For moment, the black
hole solutions in five-dimensions ($5D$)  have been considered in varius research, inspired by the ideas in string theory, braneworld cosmology and gauge/gravity duality. Certain interesting and surprising results can be found \cite{si23}. Extra dimensions voilate uniqueness theorem  due to the fact that there exist spurious degrees of freedom. The study of black-ring solutions in five-dimensions demonstrate  that higher dimensional spacetime can approve the non-trivial topologies \cite{si24}. The structure of $5D$ black hole shadows     in the context of holography were invesigated in \cite{si25}. Recently, the shadow of five-dimensional Gauss-Bonnet black hole is studied in \cite{s12}. Shadow of five-dimensional rotating Myers-Perry black hole \cite{si26, si26a} and rotating five-dimensional charged  black holes \cite{si27} have also been studied. Such investigation indeed provides a good inspiration to study the black hole shadows for RN AdS$_5$ black hole. RN AdS black hole solution in five-dimensions is a static, spherically symmetric  vacuum solution of the Einstein equation  in RN  AdS$_{5}$ spacetime. Geodesic Motions near the   RN  AdS$_{5}$ black hole are studied in \cite{s1a}. The goal of this paper is to investigate the shadow of charged  RN AdS$_{5}$ black hole in both non-plasma  and plasma medium.

This work is presented systematically in following manner. In section \ref{sec1}, we provide an overview of five-dimensional RN-AdS$_{5}$ black hole solution. From the positive real zero of horizon function, the radius of the event horizon is calculated in section \ref{sec2}.  In
section \ref{sec3}, using the Hamilton-Jacobi equation, we discuss the (null) particle  motion around the five-dimensional RN-AdS black hole in non-plasma media and eventually derive the radius of photon sphere for non-plasma medium. The shadow of this black hole for non-plasma medium is studied in section \ref{sec4}. Within section, we also tabulated the values of shadow radius and photon sphere radius for different charges. The shadow images are aso given for different values of parameters.   We study the effect of various parameters on shadow radius for non-plasma medium in section \ref{sec5}. Similar to case the non-plasma medium, geodesics in the presence of plasma medium are studied in section \ref{sec6}. The shadow for such plasma medium is presented in section \ref{sec7}. The behavior of shadow radius for various parameters in plasma medium is discussed in full details  in section \ref{sec8}. We discuss the energy emission rate of five-dimensional RN AdS black hole in section \ref{sec9}.
We see that the emission rate of RN AdS  black hole also depends on various parameters like temperature, frequency and cross-section of photon radius. Finally, we conclude our results and make future remarks  in section \ref{sec10}.

\section{$5D$  RN AdS black hole}\label{sec1}
Let us consider a $5D$ spacetime with a negative cosmological constant. The field equations
near the black hole are given by following equation \cite{s1a}:
\begin{equation}\label{1}
\bar{G}_{ab}=-\Lambda \bar{g}_{ab}+k^2_{(5)}\bar{T}_{ab},
\end{equation}
where $\bar{g}_{ab}$ is a $5D$ metric with signature $(-, +, +, +, +)$, $\bar{G}_{ab}$ denotes the $5D$ Einstein tensor, $\bar{T}_{ab}$ denotes  $5D$ energy-momentum tensor   and $\Lambda $ represents the  cosmological constant. The constant $k_{(5)}$ is defined as
\begin{equation}\label{2}
k^2_{(5)}=8\pi G =M^{-3},
\end{equation}
where $G $ is the $5D$ Newton's constant  and $M$ is the $5D$ reduced Planck mass.
 
Now, we assume that the spacetime has a constant curvature $\bar{K}=\frac{\alpha}{l^2}$. For AdS geometry, $\bar{K}$ is negative i.e, $\alpha = -1$  while  for dS geometry $\bar{K}$ becomes positive and $\alpha = 1$. Consequently, the radius of curvature of spacetime becomes
\begin{equation}\label{3}
l=\sqrt{\frac{3\alpha}{\Lambda }},
\end{equation}
and it gives the necessary length scale to have a horizon.  The exterior metric for the black hole field in $5D$ is given by
\begin{equation}\label{4}
ds^2=-f(r)dt^2+\frac{dr^2}{f(r)}+r^2d\Omega^2_{3},
\end{equation}
where the metric on unit $3$-sphere is
\begin{equation}\label{5}
d\Omega^2_{3}=d\theta^2+\sin^2\theta(d\phi^2+\sin^2\phi d\psi^2).
\end{equation}
 For this static  spherically symmetric  vacuum solution of the Einstein equations in RN AdS$_{5}$, the metric function $ f(r)$ is defined as
\begin{equation}\label{6}
f(r)=1-\left(\frac{2M}{r}\right)^2+ \frac{q^4}{r^4} -\frac{\Lambda r^2}{6}.
\end{equation}
Here, $q$, $M$ and $\Lambda$  are the total charge, geometric mass   and the cosmological constant, respectively.

\section{Horizons}\label{sec2}
We can write the lapse function (\ref{6}) as
\begin{equation}\label{7}
f(r)=\frac{\Delta}{r^4},
\end{equation}
where $\Delta$ is the horizon function that depends on $r$ only  for the given $q$, $M$   and $\Lambda$. Now, the exterior metric (\ref{4}) can be written as
\begin{equation}\label{8}
ds^2=-\frac{\Delta}{r^4}dt^2+\frac{r^4 }{\Delta}dr^2+r^2 d\theta^2+r^2\sin^2\theta(d\phi^2+\sin^2\phi d\psi^2).
\end{equation}
We know that the spacetime is singular    intrinsically
 at $r = 0$ and the nature of this singularity depends on the parameters such as cosmological constant $\Lambda$ and charge $q$. Both of these parameters can be chosen so that  spacelike naked singularity of the spacetime can be avoided \cite{car}. The metric function vanishes at the real  positive zeros of the horizon function, i.e, $\Delta = 0$, which gives
\begin{equation}\label{9}
\Lambda r^6-6r^4+24M^2r^2-6q^4=0.
\end{equation}
So, the real  positive zeros of $\Delta$ gives the position of the horizon, which indicates the coordinate singularities for this RN-AdS$_{5}$ black hole. Interestingly, among the six distinct roots of Eq. (\ref{9}), only one is  positive real, which describes  the radius of the event horizon of the RN AdS$_{5}$ black hole.  This single positive real root is given   numerically by   
\begin{equation}\label{10}
r_+=\sqrt{\frac{2}{\Lambda}-\frac{242^{\frac{1}{3}}M^2}{\mathcal{P}}+\frac{122^{\frac{1}{3}}}{\Lambda\mathcal{P}}+\frac{\mathcal{P}}{32^{\frac{1}{3}}\Lambda}},
\end{equation}
where
$$\mathcal{P} = \left(432-1296M^2\Lambda+162q^4\Lambda^2+\sqrt{4(-36+72M^2\Lambda)^3+(432-1296M^2\Lambda+162q^4\Lambda^2)^2}\right)^{\frac{1}{3}}.$$

\section{Particle motion for non-plasma medium}\label{sec3}
In order to find the shape of the  RN AdS$_{5}$ black hole  shadow, we first require to derive the geodesic equations for the photons (null particle) around the black holes. The directions of symmetry  basically work on the issue of searching the geodesics by knowing the constants of motion related to the directions of symmetries. Let us take $k^{\mu}$ as a vector towards the direction of symmetry and $u^{\mu}=\frac{d x^{\mu}}{d \lambda}$  along perpendicular   direction tangent on a curve $x^{\mu}=x^{\mu}(\lambda)$ with affine parameter $\lambda$. Now, with the help of the Killing equation, it can be shown that
\begin{equation}\label{11}
k^{\mu} u_{\mu}= \mbox{constant},
\end{equation}
when the trajectory $x^{\mu}$ is a geodesic \cite{s1}. As the coefficients of  metric (\ref{4}) are independent of time, this results  a timelike Killing vector. Therefore,  Eq. (\ref{11}) becomes
\begin{equation}\label{12}
k^{0} u_{0}=u_{0}=-E,
\end{equation}
where constant $E$ is known as the relativistic energy per particle mass.
The negative sign in front of  $E$ is chosen for convenience only and has no other physical significance. In fact, coefficients of the metric  are  independent of $\phi$ and $\psi$ as well, so these coordinates  represent  the other symmetry directions. Hence, by choosing $k^{\mu}=(0,0,0,1,0)$ along the $\phi$ direction,  we get 
\begin{equation}\label{13}
k^{\mu} u_{\mu}=u_{3}=L_{\phi},
\end{equation}
and $k^{\mu}=(0,0,0,0,1)$ along the $\psi$ direction, we get
\begin{equation}\label{14}
k^{\mu} u_{\mu}=u_{4}=L_{\psi},
\end{equation}
where constants $L_{\phi}$ and $L_{\psi}$ are known as the angular momentum per particle mass.

 Now, the geodesic equations along  the $t$, $\phi$ and $\psi$ directions can be calculated by utilizing the above constants of motion. These read
\begin{eqnarray}
\frac{d t}{d \lambda} &=&\frac{E}{f(r)},\label{15}\\
\frac{d \phi}{d \lambda} &=&\frac{L_{\phi}}{r^{2} \sin ^{2} \theta},\label{16}\\
\frac{d \psi}{d \lambda} &=&\frac{L_{\psi}}{r^{2} \cos ^{2} \theta}.\label{17}
\end{eqnarray}
The remaining two geodesic equations can be calculated with the help of the relativistic Hamilton-Jacobi equation given as
\begin{equation}\label{18}
\frac{\partial S}{\partial \lambda}+\frac{1}{2} g^{\mu \sigma} \frac{\partial S}{\partial x^{\mu}} \frac{\partial S}{\partial x^{\sigma}}=0.
\end{equation}
Now, in order to solve the above
 Hamilton-Jacobi equation, we take an ansatz for $S$ of the form  \cite{si8}
\begin{equation}\label{19}
S=\frac{1}{2} m_{0}^{2} \lambda-E t+L_{\phi} \phi+L_{\psi} \psi+H_{r}(r)+H_{\theta}(\theta),
\end{equation}
where $H_{r}(r)$ and $H_{\theta}(\theta)$ are functions of only $r$ and $\theta$,  respectively. Here, $\lambda$  and $m_{0}$  signify  the affine parameter and the rest mass of the test particle, respectively. Using Eqs. (\ref{19}) and (\ref{18}), we obtain
\begin{equation}\label{20}
\left(\frac{\partial H_{\theta}}{\partial \theta}\right)^{2}+L_{\phi}^{2} \cot ^{2} \theta+L_{\psi}^{2} \tan ^{2} \theta+\frac{1}{2} m_{0}^{2}-\frac{r^{2} E^{2}}{f(r)}+r^{2} f(r)\left(\frac{\partial H_{r}}{\partial r}\right)^{2}+L_{\phi}^{2}+L_{\psi}^{2}=0.
\end{equation}
This further simplifies to
\begin{equation}\label{21}
\left(\frac{\partial H_{\theta}}{\partial \theta}\right)^{2}+L_{\phi}^{2} \cot ^{2} \theta+L_{\psi}^{2} \tan ^{2} \theta+\frac{1}{2} m_{0}^{2}=\frac{r^{2} E^{2}}{f(r)}-r^{2} f(r)\left(\frac{\partial H_{r}}{\partial r}\right)^{2}-L_{\phi}^{2}-L_{\psi}^{2}=\mathcal{C}
\end{equation}
where $\mathcal{C}$ is the Carter constant. As we know the relation $p_{\theta}=\frac{\partial H}{\partial \theta}=\frac{\partial H_{\theta}}{\partial \theta}$, so we have
\begin{equation}\label{22}
\frac{\partial H_{\theta}}{\partial \theta}=r^{2} \frac{\partial \theta}{\partial \lambda}.
\end{equation}
Here, the  condition $p_{\theta}=\frac{\partial \mathcal{L}}{\partial \theta}$ 
is used and Lagrangian $\mathcal{L}$ is given by
\begin{equation}\label{23}
\mathcal{L}=\frac{1}{2} g_{\mu \nu} \frac{d{x}^{\mu}}{d\lambda} \frac{d{x}^{\nu}}{d\lambda}.
\end{equation}
  Similarly, using the condition $p_{r}=\frac{\partial H}{\partial r}=\frac{\partial H_{r}}{\partial r}$, we can write
\begin{equation}\label{24}
\frac{\partial H_{r}}{\partial r}=r^{2} \frac{\partial r}{\partial \lambda}.
\end{equation}
In order to determine the null geodesics, we set $m_{0}=0$ and plugging  the values of the Eqs. (\ref{22} and (\ref{24}) into Eq.(\ref{21}), we obtain
\begin{equation}\label{25}
r^{2}\left(\frac{d r}{d \lambda}\right)=\sqrt{r^{4} E^{2}-(L^{2}+\mathcal{C}) r^{2} f(r)},
\end{equation}
\begin{equation}\label{26}
r^{2}\left(\frac{d \theta}{d \lambda}\right)=\sqrt{\mathcal{C} -L_{\phi}^{2} \cot ^{2} \theta-L_{\psi}^{2} \tan ^{2} \theta},
\end{equation}
 where  $L^{2}= L_{\phi}^{2}+L_{\psi}^{2}$. The above Eqs. (\ref{25}) and (\ref{26}) are the geodesic equations along $r$ and $\theta$, respectively.
 
Now, using Eq. (\ref{25}), we can arrive at the   the familiar form of equation:
\begin{equation}\label{29}
\left(\frac{d r}{d \lambda}\right)^{2}+V (r)=0,
\end{equation}
where $V $ is an effective potential, given by
\begin{eqnarray}\label{30}
V(r)&=&\frac{f(r)}{r^{2}}\left(\mathcal{C} +L^{2}\right)-E^{2}.
\end{eqnarray}
To find the unstable circular orbits, we consider the following conditions:
\begin{equation}\label{31}
\left.V (r)\right|_{r=r_{p}}=0,\;\left.\frac{\partial V (r)}{\partial r}\right|_{r=r_{p}}=0.
\end{equation}
Now, $V (r)$ will be a maximum at $r=r_{p}$ when
\begin{equation}\label{32}
\left.\frac{\partial^{2} V (r)}{\partial r^{2}}\right|_{r=r_{p}}<0,
\end{equation}
where $r_{p}$ is the photon sphere radius. 
Now, using Eq. (\ref{30}), the first condition of (\ref{31}) leads to
\begin{equation}\label{33}
\frac{r_{p}^{2}}{ f(r_{p})} = 
 \eta+\xi^{2},\;\; \mbox{with}\ \  \xi^{2} \equiv \xi_{1}^{2}+\xi_{2}^{2},
\end{equation}
where   the Chandrasekhar constants $\eta, \xi_{1}$ and $\xi_{2}$ are considered to have forms \cite{si8}  
\begin{equation}\label{34}
\eta=\frac{\mathcal{C}}{E^{2}},\; \quad \xi_{1}=\frac{L_{\phi}}{E},\;\;\;\xi_{2}=\frac{L_{\psi}}{E}.
\end{equation}
Now, the (boundary) condition, $\left.\frac{\partial V_{e f f}(r)}{\partial r}\right|_{r=r_{p}}=0$, yields
\begin{equation}\label{35}
\left.r \frac{df(r)}{dr}  \right|_{r=r_{p}}=\left. 2 f(r)\right|_{r=r_{p}}.
\end{equation}
For the given metric function $f(r)$ in (\ref{6}),  the condition (\ref{35}) leads to
\begin{equation}\label{37}
r^4_{p}-8M^2r^2_{p}+3q^4=0.
\end{equation}
The solution of this equation results to the radius of the photon sphere as
\begin{equation}\label{38}
r_{p}=\sqrt{\frac{8M^2+\sqrt{64M^4-12q^4}}{2}}.
\end{equation}

\section{Constructing the black hole shadow}\label{sec4}
\begin{figure}[ht]
\begin{center}
\includegraphics[width=0.5\linewidth]{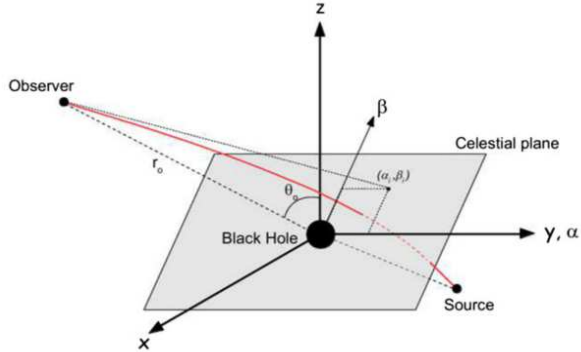} 
\end{center}
\caption{Schematic diagram of celestial coordinates. Courtesy: \cite{s4}.}
\label{fig1}
\end{figure}

To get the shadow of the RN-AdS$_{5}$ black hole, we   introduce the celestial coordinates as shown schematically in figure \ref{fig1}. For $5D$, the celestial coordinates become  \cite{s3}
\begin{eqnarray}\label{39}
\alpha &=&\lim _{r \rightarrow \infty} \left(r^{2} \sin \theta \frac{d \phi}{d r}+r^{2} \cos \theta \frac{d \psi}{d r}\right),\nonumber\\
&\beta& =\lim _{r \rightarrow \infty} r^{2} \sin \theta \frac{d \theta}{d r},
\end{eqnarray}
where $\alpha$ component    represents  the apparent perpendicular displacement of the shadow as observed from the $z$-axis (axis of symmetry) and $\beta$ component represents the the apparent perpendicular displacement of the shadow as observed from its projection on the equatorial surface. Here, $r_{0}$ denotes the distance between the viewer and the black hole. However,  $\theta_0$ is the angle between the observer's eye line and the axis of rotation of the black hole called as   the inclination angle. Recalling the geodesic equations derived in Eqs. (\ref{15}) and (\ref{25}), we get the values of $\frac{d \phi}{d r}, \frac{d \psi}{d r}$ and $\frac{d \theta}{d r}$ as following:
\begin{equation}\label{40}
\frac{d \phi}{d r} =\frac{L_{\phi} \csc ^{2} \theta}{r^{2} \sqrt{E^{2}-\frac{f(r)}{r^{2}}\left(\mathcal{C}+L_{\phi}^{2}+L_{\psi}^{2}\right)}},
\end{equation}
\begin{equation}\label{41}
\frac{d \psi}{d r} =\frac{L_{\phi} \sec ^{2} \theta}{r^{2} \sqrt{E^{2}-\frac{f(r)}{r^{2}}\left(\mathcal{C}+L_{\phi}^{2}+L_{\psi}^{2}\right)}},
\end{equation}
\begin{equation}\label{42}
\frac{d \theta}{d r} =\frac{1}{r^{2}} \sqrt{\frac{\mathcal{C}-L_{\phi}^{2} \cot ^{2} \theta-L_{\psi}^{2} \tan ^{2} \theta}{E^{2}-\frac{f(r)}{r^{2}}\left(\mathcal{C}+L_{\phi}^{2}+L_{\psi}^{2}\right)}}.
\end{equation}
Plugging the these equations into the Eq. (\ref{39}), we get
\begin{equation}\label{43}
\alpha=-\frac{ \xi_{1} \csc \theta+\xi_{2} \sec \theta }{\sqrt{1+\left(\eta+\xi_{1}^{2}+\xi_{2}^{2}\right)\Lambda}},
\end{equation}
\begin{equation}\label{44}
\beta=\pm \sqrt{\frac{ \eta-\xi_{1}^{2} \cot ^{2} \theta-\xi_{2}^{2} \tan ^{2} \theta }{1+\left(\eta+\xi_{1}^{2}+\xi_{2}^{2}\right)\Lambda}}.
\end{equation}
Let us consider two different values for $\theta$, i.e., $0$ and $ \frac{\pi}{2}$. If $\theta=\frac{\pi}{2}$ then $L_{\psi}=0$ and as a result $\xi_{1} \equiv \xi$. However, when $\theta=0$, $L_{\phi}=0$ which indicates $\xi_{2} \equiv \xi$. Taking both the cases together into account, the above celestial coordinates take  the form
\begin{equation}\label{45}
\alpha=-\frac{\xi}{\sqrt{1+\left(\eta+\xi^{2}\right)\Lambda}},
\end{equation}
\begin{equation}\label{46}
\beta=\pm \sqrt{\frac{\eta}{1+\left(\eta+\xi^{2}\right)\Lambda}}.
\end{equation}
Combining the eqs. (\ref{45}) and (\ref{46}), we arrive at the equation of a circle in the  $\alpha\beta$-plane 
\begin{equation}\label{47}
\alpha^{2}+\beta^{2}=\frac{\eta+\xi^{2}}{1+(\eta+\xi^{2})\Lambda} \equiv R_{s}^{2} 
\end{equation}
where $R_{s}$ refers to the radius of the shadow for non-plasma medium.
 
We have shown the numerical values of shadow radius   and photon sphere radius   of RN-AdS$_{5}$ black hole for different values   $q$ and cosmological constant $\Lambda$ in table \ref{table:1} and table \ref{table:2}. 
\begin{table}[h!]
\centering
\begin{tabular}{||c c   c||} 
 \hline
 $q$ & $r_{p}$ &   $R_{s}$ \\ [0.5ex] 
 \hline\hline
  0.1& 2.82842 &   3.76177 \\ 
  0.3& 2.82789 &  3.76135 \\
 0.5 & 2.82427 &   3.75851 \\
  0.7&2.81228  &   3.74914 \\[1ex]
 \hline 
\end{tabular}
\caption{Photon radius ($r_{p}$) and the black hole shadow  radius  ($R_{s}$) for varying $q$ with $\Lambda = 0.0098$.}
\label{table:1}
\end{table}
\begin{table}[h!]
\centering
\begin{tabular}{||c c   c||} 
 \hline
 $q$ & $r_{p}$ &   $R_{s}$ \\ [0.5ex] 
 \hline\hline
  0.1& 2.82842 &    3.5921 \\ 
  0.3& 2.82789 &    3.59174 \\
 0.5 & 2.82427 &   3.58927 \\
  0.7&2.81228  &   3.5811 \\[1ex]
 \hline
\end{tabular}
\caption{Photon radius ($r_{p}$) and the black hole shadow  radius ($R_{s}$) for varying $q$ with $\Lambda = 0.018$.}
\label{table:2}
\end{table}

\begin{figure}[h!]
\begin{center} 
$\begin{array}{cccc}
\subfigure[]{\includegraphics[width=0.5\linewidth]{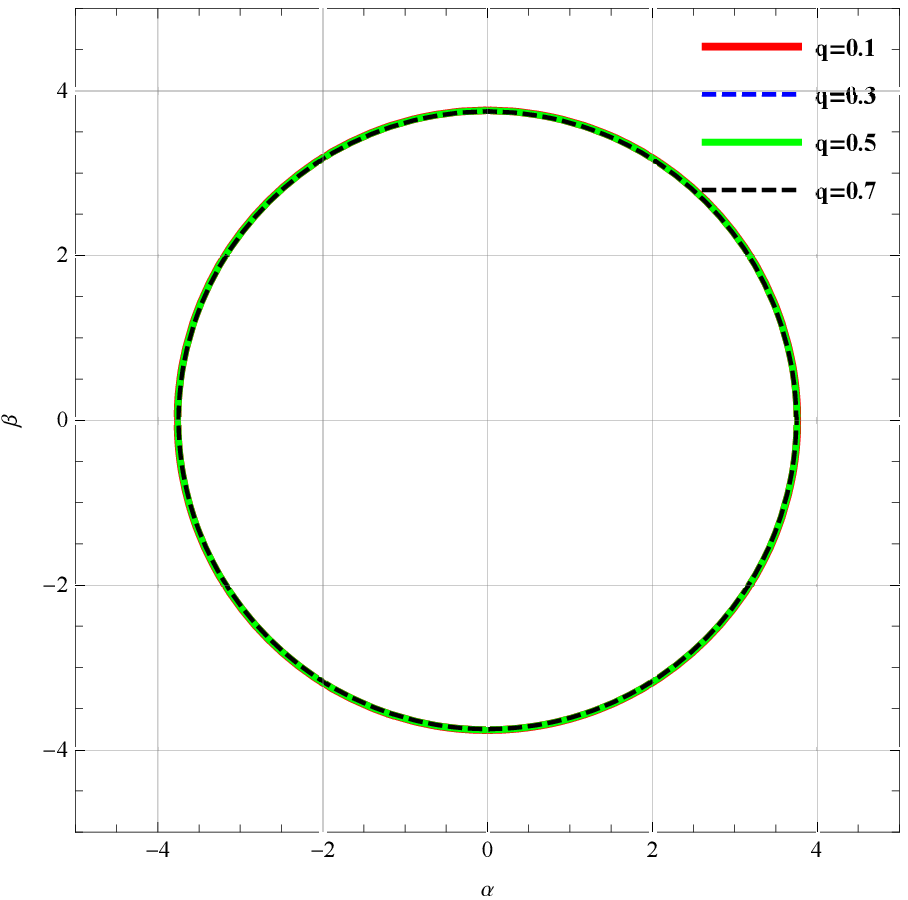}
\label{2a}}
\subfigure[]{\includegraphics[width=0.5\linewidth]{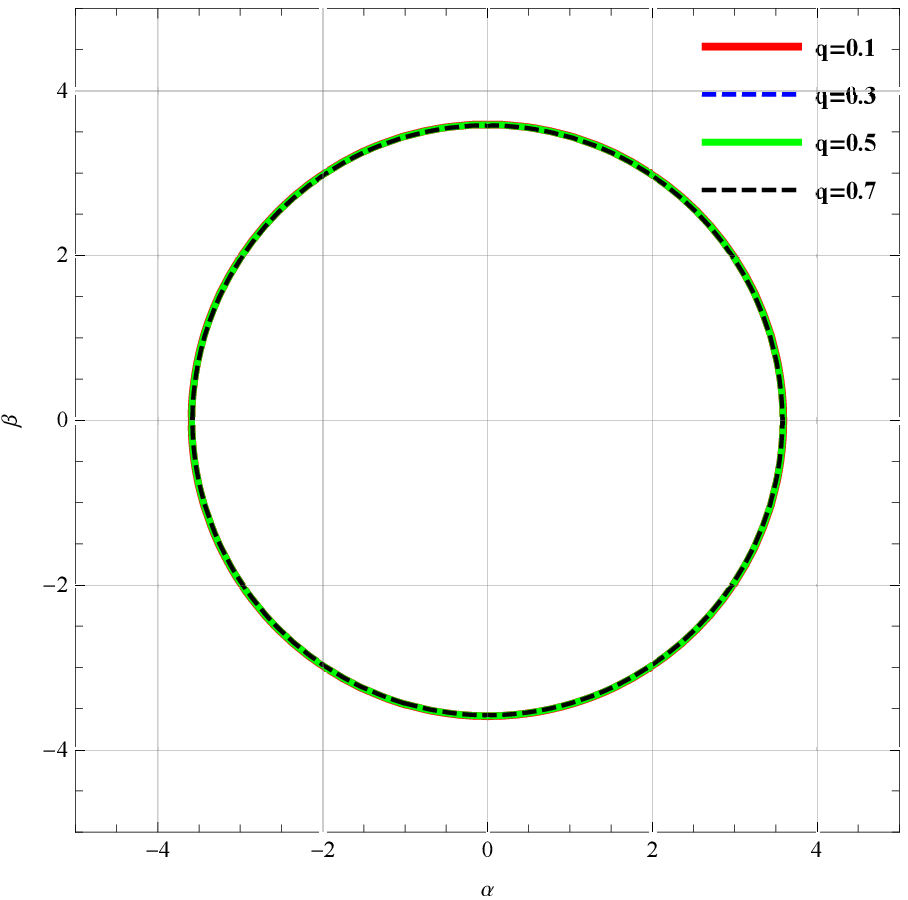}\label{2b}} 
\end{array}$
\end{center}
\caption{In (a), black hole shadow in  $\alpha \beta$-plane  for different values of $q$ with a cosmological constant $\Lambda = 0.0098$. In (b), black hole shadow in  $\alpha \beta$-plane  for different values of $q$ with cosmological constant $\Lambda = 0.018$. Here, $M = 1$.}
\label{fig2}
\end{figure}

In figure \ref{fig2}, the variation of the black hole shadow in celestial plane   for different values of the charge $q$ and cosmological constant $\Lambda=0.0098,0.018$ is shown graphically. 
Here, we observe  that the radius of the black hole shadow is a decreasing function of charge parameter $q$ and  cosmological constant $\Lambda$.

\section{Effect of  parameters on shadow radius in non-plasma medium}\label{sec5}
Here, we study the effect of various parameter on shadow radius $R_{s}$ for non-plasma medium. Now,
   the shadow radius in non-plasma medium  takes the following form:
\begin{equation}\label{48}
R_{s}=\sqrt{\frac{\eta+\xi^{2}}{1+(\eta+\xi^{2})\Lambda}}=\sqrt{ \frac{r^2_{p}}{ f(r_{p})+\Lambda {r^2_{p}} }}.
\end{equation}
Here, we have used relation (\ref{33}). It is noticed that shadow radius depends on the parameters like cosmological constant $\Lambda$, charge $q$ and mass $M$ of the black hole.
\begin{figure}[h!]
\begin{center} 
$\begin{array}{cccc}
\subfigure[]{\includegraphics[width=0.5\linewidth]{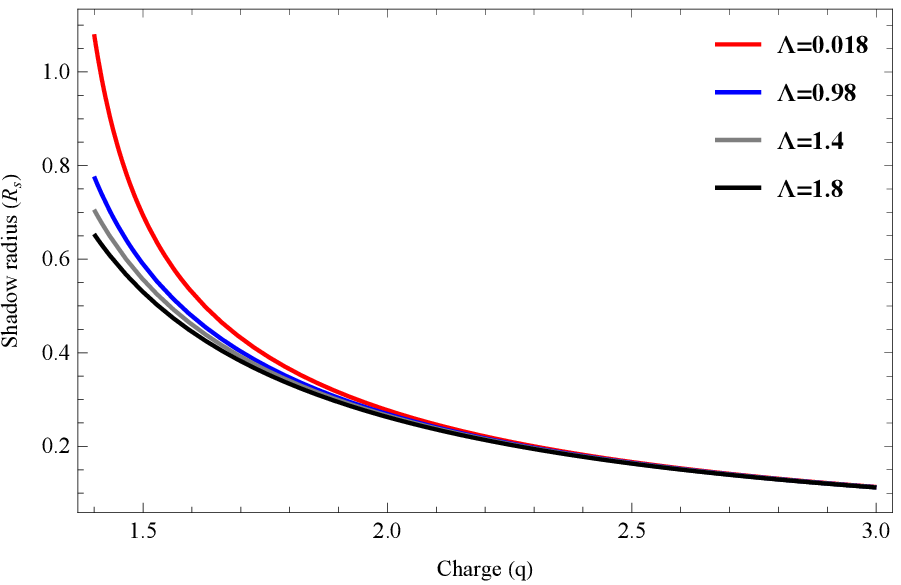}
\label{3a}}
\subfigure[]{\includegraphics[width=0.5\linewidth]{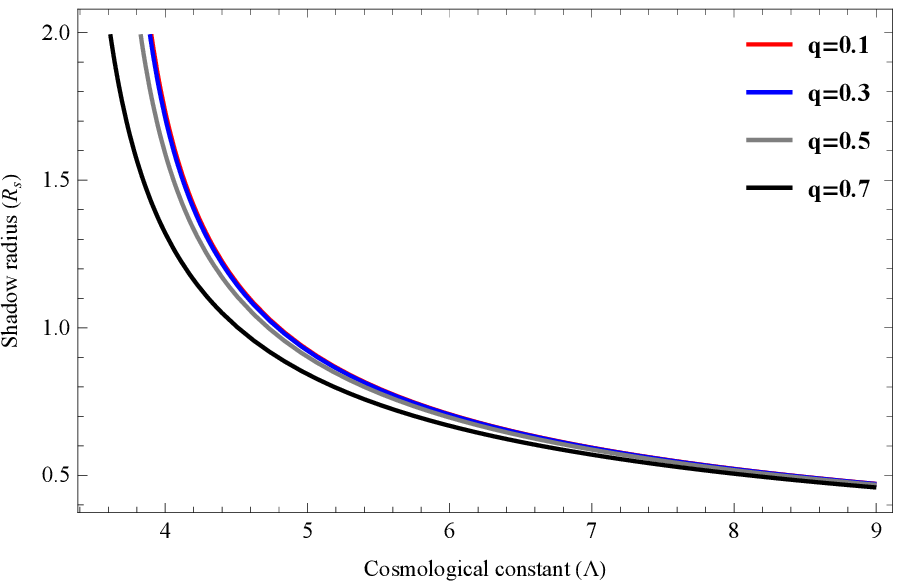}\label{3b}} 
\end{array}$
\end{center}
\caption{In (a), the black hole shadow radius  vs.  charge   for changing cosmological constant $\Lambda$. In (b),  the black hole shadow radius  vs. cosmological constant for changing charge $q$. Here, $M = 1$.}
\label{fig3}
\end{figure}

Figure \ref{fig3} depicts how shadow radius for non-plasma medium depends on parameter like charge $q$ and cosmological constant $\Lambda$. In \ref{3a}, shadow radius is a decreasing function of charge for fixed photon sphere radius. We see that, for larger  values of $\Lambda$, shadow radius curve decreases sharply. However, in plot \ref{3b}, shadow radius is a decreasing function of $\Lambda$ and here shadow radius curve decreases for increasing charge parameter $q$. 
\section{Shadow cast in the presence of  plasma medium}\label{sec6}
 Generally, a black hole is enclosed by a   substance media which gets affected on the motion of null like particles   passing through it.
In the present section, we emphasize the consequences  of a plasma medium on the shadow of RN-AdS$_{5}$ black hole. For the plasma medium , the refractive index    is denoted by $n=n\left(x^{i}, \omega\right)$, where $\omega$ is the angular frequency of photon  measured by a distant  moving viewer having a velocity $u^{\mu}$. The Hamiltonian is now modified by this plasma background. Due to this medium, there appears an extra term  in the equations of geodesic motion and, hence, the trajectories of null paricles get modified.  Due to the plasma effect, effective energy of the particle also gets modified and  given by $E=$ $\hbar \omega=-p_{\alpha} u^{\alpha}$. The relationship between the refractive index and the $4$-momentum of the photon is given by \cite{s5}
\begin{equation}\label{49}
n^{2}=1+\frac{p_{\alpha} u^{\alpha}}{\left(p_{\mu} u^{\mu}\right)^{2}}.
\end{equation}
Now,  the refractive index ($n$) and plasma frequency ($\omega_{p}$) arw related by \cite{s6}
\begin{equation}\label{50}
n=\sqrt{1-\left(\frac{\omega_{p}}{\omega}\right)^{2}},
\end{equation}
where $\omega_{p}$ is defined by
\begin{equation}\label{51}
\omega_{p}=\frac{4 \pi e^{2} N }{m_{e}}.
\end{equation}
Here, $e, N $ and $m_{e}$ are, respectively, the charge, $r$ dependent number density and electron mass in the plasma medium. The physical form of $N(r)$ is considered as $\frac{N_{0}}{r^{i}}$  \cite{s5, s6}, where $i$ identifies various characteristics of the plasma medium but we consider $i=1$ that reflects the minimum dependency on $r$ \cite{s6, s7}.   Replacing the form of $N(r)$ in the plasma frequency $\omega_{p}$ and, from Eq.(\ref{50}), we have  
\begin{equation}\label{52}
\left(\frac{\omega_{p}}{\omega}\right)^{2}=\frac{k}{r}, k \geqslant 0.
\end{equation}
  Consequently, the refractive index $n$ becomes
\begin{equation}\label{53}
n=\sqrt{1-\frac{k}{r}}.
\end{equation}
The corrcted  Hamilton-Jacobi equation reads \cite{s5}
\begin{equation}\label{54}
\left(\frac{\partial S}{\partial \lambda}\right)+\frac{1}{2}\left[g^{\mu \sigma} \frac{\partial S}{\partial x^{\mu}} \frac{\partial S}{\partial x^{\sigma}}-\left(n^{2}-1\right)\left(\frac{\partial S}{\partial t} \sqrt{-g^{t t}}\right)^{2}\right]=0.
\end{equation}
\section{Geodesics and the black hole shadow for the plasma medium}\label{sec7}
In order to study the consequences of a plasma medium on black hole shadow, we require  a new set of (celestial) coordinates. We begin the study by considering the modified geodesic equations. Now, for plasma medium, the set of null geodesics read
\begin{eqnarray}
\frac{d t}{d \lambda} & =&\frac{n^{2} E}{f(r)},\label{55} 
\\
\frac{d \phi}{d \lambda} & =&\frac{L_{\phi}}{r^{2} \sin ^{2} \theta},\label{56}\\
\frac{d \psi}{d \lambda} & =&\frac{L_{\psi}}{r^{2} \cos ^{2} \theta},\label{57}\\
r^{2}\left(\frac{d r}{d \lambda}\right)& =&\pm \sqrt{n^{2} r^{4} E^{2}-\left(L^{2}+\mathcal{C}\right) r^{2} f(r)},\label{58}
\\
r^{2}\left(\frac{d \theta}{d \lambda}\right) & =&\pm \sqrt{\mathcal{C}-L_{\phi}^{2} \cot ^{2} \theta-L_{\psi}^{2} \tan ^{2} \theta}.\label{59}
\end{eqnarray}
Following the same  procedure as in the non-plasma medium, the effective radial potential in   the plasma medium becomes
\begin{equation}\label{62}
V^{p l}(r)=\frac{f(r)}{r^{2}}\left(\mathcal{C}+L^{2}\right)-n^{2} E^{2}.
\end{equation}
The unstable circular orbit can be determined by
\begin{equation}\label{63}
\left. V^{p l}(r)\right|_{r=r_{p}^{(p l)}}=0,\;\;\left.\frac{\partial V^{p l}(r)}{\partial r}\right|_{r=r_{p}^{(p l)}}=0,
\end{equation}
where $r_{p}^{(p l)}$ denotes radius of  photon sphere in plasma medium.
The condition for maximizing   $V^{p l}(r)$ reads
\begin{equation}\label{64}
\left.\frac{\partial^{2} V^{p l}(r)}{\partial r^{2}}\right|_{r=r_{p}^{(p l)}}<0.
\end{equation}
The first condition of Eq. (\ref{63}) leads to
\begin{equation}\label{65}
\eta+\xi^{2}=\left.\frac{n^{2}  r^{2}}{f }\right|_{r=r_{p}^{(p l)}},
\end{equation}
and the second condition gives
\begin{equation}\label{66}
\left.\left(n  r f^{\prime} -2 n  f -2 n^{\prime}  r f \right)\right|_{r=r_{p}^{(p l)}}=0.
\end{equation}

Now, substituting  the values of $f(r)$   (\ref{6})  and refractive index $n^{\prime}(r)$    (\ref{53})) in (\ref{66}), we obtain the equation for the radius of the photon sphere as
\begin{eqnarray}\label{67}
 \sqrt{1-\frac{k}{r_{p}}}\left[ \frac{4M^2}{r^2_{p}} -\frac{6q^4}{r^4_{p}}-2\right] -\frac{k}{r_{p}\sqrt{1-\frac{k}{r_{p}}}}\left[1- \frac{4M^2}{r_{p}^2} + \frac{q^4}{r^4_{p}} -\frac{\Lambda r^2_{p}}{6}\right]=0.
\end{eqnarray}
Here, we notice that an exact solution of Eq. (\ref{67}) is not possible. So, we try to solve this equation numerically. Due to the  plasma medium,  an additional parameter ($k$) arises in Eq. (\ref{67}) . Let us fix $k$ to take  two values, namely, $0.2$ and $0.4$ and for these values we obtain the radius of photon   $r_{p}$ upon solving  (\ref{67}) numerically. Following the same procedures as before, we obtain an expression  for $\frac{d \phi}{d r}, \frac{d \psi}{d r}$ and $\frac{d \theta}{d r}$ for the plasma medium and calculate the celestial coordinates $(\alpha, \beta)$ for this medium. These  are
\begin{equation}\label{68}
\frac{d \phi}{d r}=\frac{L_{\phi} \csc ^{2} \theta}{r^{2} \sqrt{n^{2} E^{2}-\frac{f(r)}{r^{2}}\left(\mathcal{C}+L_{\phi}^{2}+L_{\psi}^{2}\right)}},
\end{equation}
\begin{equation}\label{69}
\frac{d \psi}{d r}=\frac{L_{\phi} \sec ^{2} \theta}{r^{2} \sqrt{n^{2} E^{2}-\frac{f(r)}{r^{2}}\left(\mathcal{C}+L_{\phi}^{2}+L_{\psi}^{2}\right)}},
\end{equation}
\begin{equation}\label{70}
\frac{d \theta}{d r}=\frac{1}{r^{2}} \sqrt{\frac{\mathcal{C}-L_{\phi}^{2} \cot ^{2} \theta-L_{\psi}^{2} \tan ^{2} \theta}{n^{2} E^{2}-\frac{f(r)}{r^{2}}\left(\mathcal{C}+L_{\phi}^{2}+L_{\psi}^{2}\right)}}.
\end{equation}
Plugging the above values into the expressions of the $(\alpha, \beta)$ coordinates   (\ref{39}), we get
\begin{eqnarray}
\alpha &=&-\frac{\left(\xi_{1} \csc \theta+\xi_{2} \sec \theta\right)}{\sqrt{1+\left(\eta+\xi_{1}^{2}+\xi_{2}^{2}\right)\Lambda}},    \label{71}\\
\beta &=&\pm \sqrt{\frac{\Big(\eta-\xi_{1}^{2} \cot ^{2} \theta-\xi_{2}^{2} \tan ^{2} \theta\Big)}{1+\Big(\eta+\xi_{1}^{2}+\xi_{2}^{2}\Big)\Lambda}}.\label{72}
\end{eqnarray}    
Here, we fix  $\theta$ to have two different values, i.e, $\frac{\pi}{2}$ and $0$. For both these values together, the   coordinates $(\alpha, \beta)$  becomes 
\begin{equation}\label{73}
\alpha=-\frac{\xi}{\sqrt{1+\left(\eta+\xi^{2}\right)\Lambda}},
\end{equation}
\begin{equation}\label{74}
\beta=\pm \sqrt{\frac{\eta}{1+\left(\eta+\xi^{2}\right)\Lambda}}.
\end{equation}
Taking the celestial coordinates from Eqs. (\ref{73}) and (\ref{74})  into account and using Eq. (\ref{65}), we have
\begin{equation}\label{75}
\alpha^{2}+\beta^{2} = \frac{n^2r^2_{p}} {f(r_{p})+\Lambda {n^2r^2_{p}} },
\end{equation}
where $R_{s}:= \sqrt{\alpha^{2}+\beta^{2}}$ is the radius of the  shadow.
 
\begin{figure}[h!]
\begin{center} 
$\begin{array}{cccc}
\subfigure[]{\includegraphics[width=0.5\linewidth]{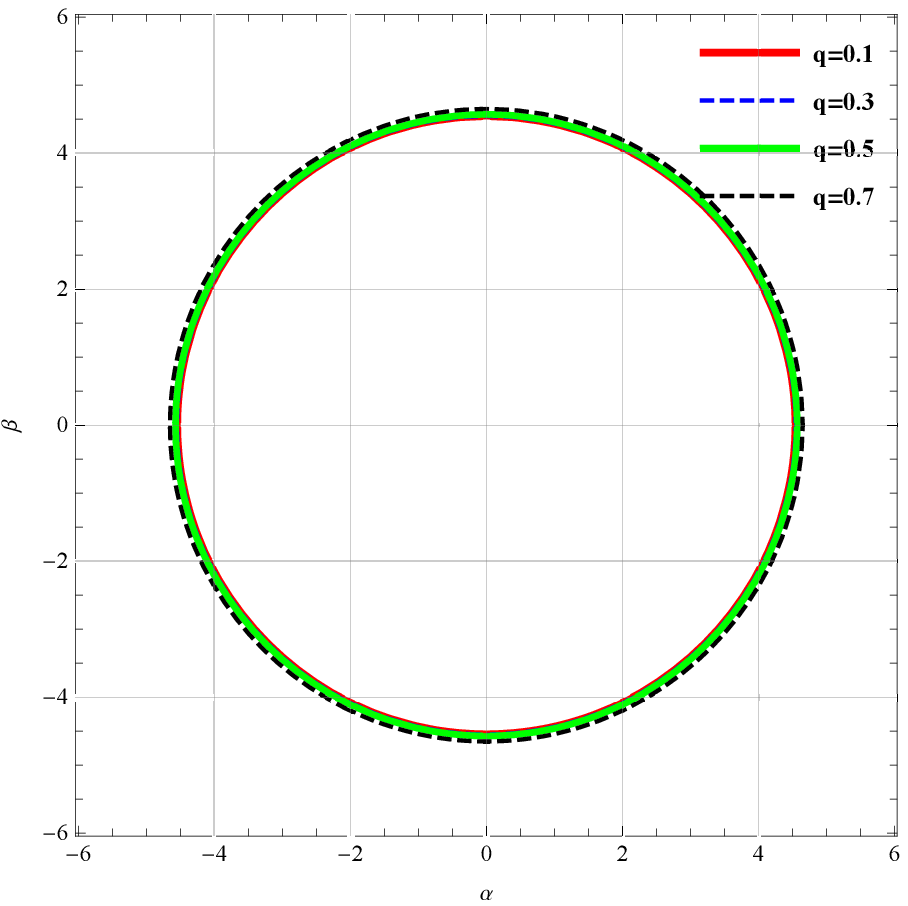}
\label{4a}}
\subfigure[]{\includegraphics[width=0.5\linewidth]{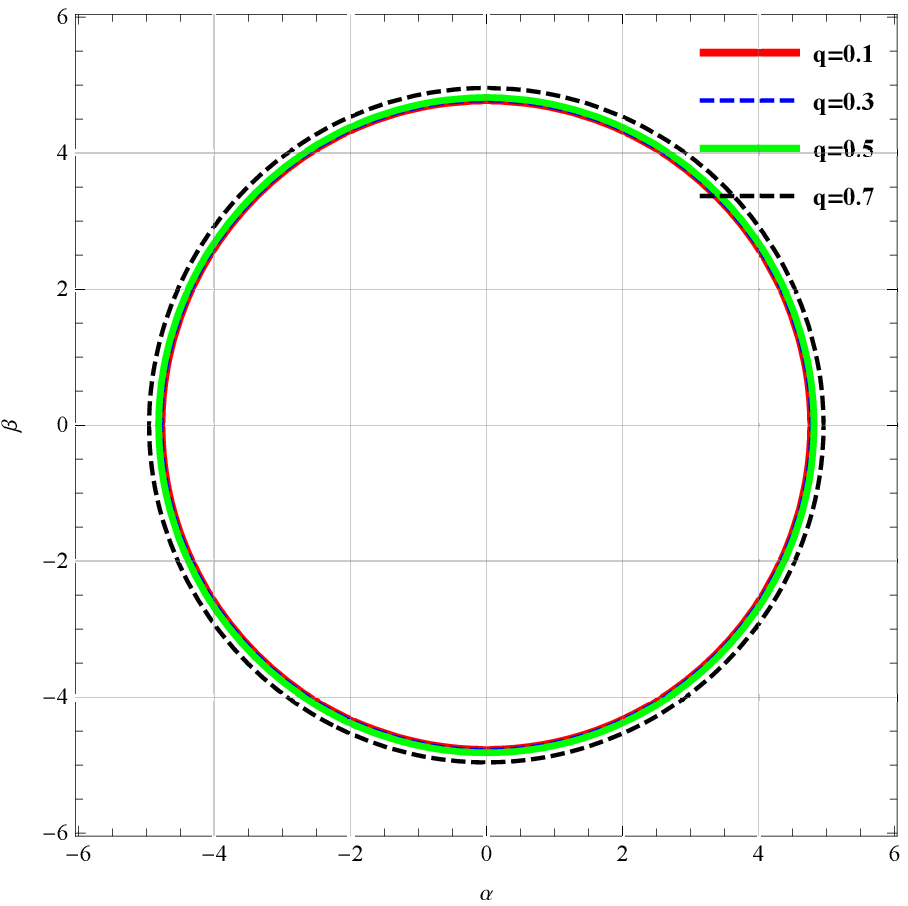}\label{4b}} 
\end{array}$
\end{center}
\caption{In  (a), black hole shadow in  $\alpha \beta$-plane for different values of $q$ with $k=0.2$. In (b), black hole shadow in the  $\alpha\beta$-plane  for different values of $q$ with $k=0.4$. Here, $\Lambda = 0.018$ and $M = 1$.}
\label{fig4}
\end{figure}

Tables \ref{table:3}, \ref{table:4}, \ref{table:5} and \ref{table:6} illustrate  the  values of  shadow radius  and photon radius  for different charge $q$ of the RN AdS$_{5}$ black hole  and cosmological constant $\Lambda$ in  the plasma medium.
\begin{table}[h!]
\centering
\begin{tabular}{||c c c ||} 
 \hline
 q & $r_{p}$ & $R_{s}$  \\ [0.5ex] 
 \hline\hline
  0.1& 2.16013 & 4.54487  \\ 
  0.3& 2.15914 & 4.54799  \\
 0.5 & 2.15233 & 4.57085  \\
  0.7& 2.12942 & 4.65152  \\[1ex]
 \hline
\end{tabular}
\caption{ $r_{p}$  and $R_{s}$ for varying $q$ with $k=0.2$, $\Lambda = 0.018$ and $M=1$.}
\label{table:3}
\end{table}
\begin{table}[h!]
\centering
\begin{tabular}{||c c  c||} 
 \hline
 q & $r_{p}$ & $R_{s}$  \\ [0.5ex] 
 \hline\hline
  0.1& 2.15985 & 4.89338  \\ 
  0.3& 2.15885 & 4.90248  \\
 0.5 & 2.15205 & 4.93012  \\
  0.7& 2.12914 & 5.03282  \\[1ex]
 \hline
\end{tabular}
\caption{ $r_{p}$  and $R_{s}$  for varying $q$ with $k=0.2$, $\Lambda = 0.0098$ and $M=1$.}
\label{table:4}
\end{table}
\begin{table}[h!]
\centering
\begin{tabular}{||c c  c||} 
 \hline
 q & $r_{p}$ & $R_{s}$  \\ [0.5ex] 
 \hline\hline
  0.1& 2.11695 & 4.77485  \\ 
  0.3& 2.11592 & 4.78018  \\
 0.5 & 2.10887 & 4.81765  \\
  0.7& 2.08512 & 4.95895 \\[1ex]
 \hline
\end{tabular}
\caption{ $r_{p}$  and $R_{s}$  for changing $q$ with $k=0.4$, $\Lambda = 0.018$ and $M=1$.}
\label{table:5}
\end{table} 
\begin{table}[h!]
\centering
\begin{tabular}{||c c  c||} 
 \hline
 q & $r_{p}$ & $R_{s}$  \\ [0.5ex] 
 \hline\hline
  0.1& 2.11636 & 5.18284  \\ 
  0.3& 2.11533 & 5.18966  \\
 0.5 & 2.10828 & 5.23775  \\
  0.7& 2.08453 & 5.421  \\[1ex]
 \hline
\end{tabular}
\caption{ $r_{p}$  and $R_{s}$  for varying $q$ with $k=0.4$, $\Lambda = 0.0098$ and $M=1$.}
\label{table:6}
\end{table}
In celestial plane, the dependence of the black hole shadow on charge $q$ and plasma parameter $k$ with cosmological constant $\Lambda = 0.018$ for plasma medium is depicted in figure \ref{fig4}. We see that shadow radius increases with plasma parameter $k$ for different photon sphere radius. We also   observe that the radius of the black hole shadow increases for an increase in the value of the charge parameter $q$.

\begin{figure}[h!]
\begin{center} 
$\begin{array}{cccc}
\subfigure[]{\includegraphics[width=0.5\linewidth]{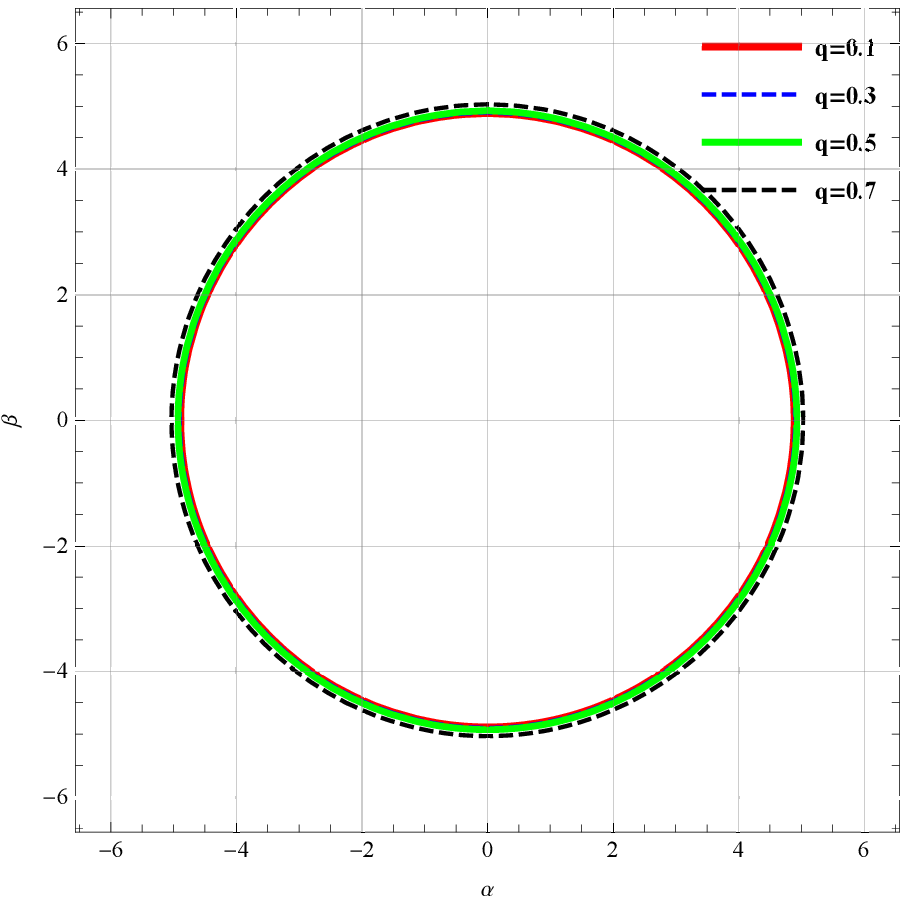}
\label{5a}}
\subfigure[]{\includegraphics[width=0.5\linewidth]{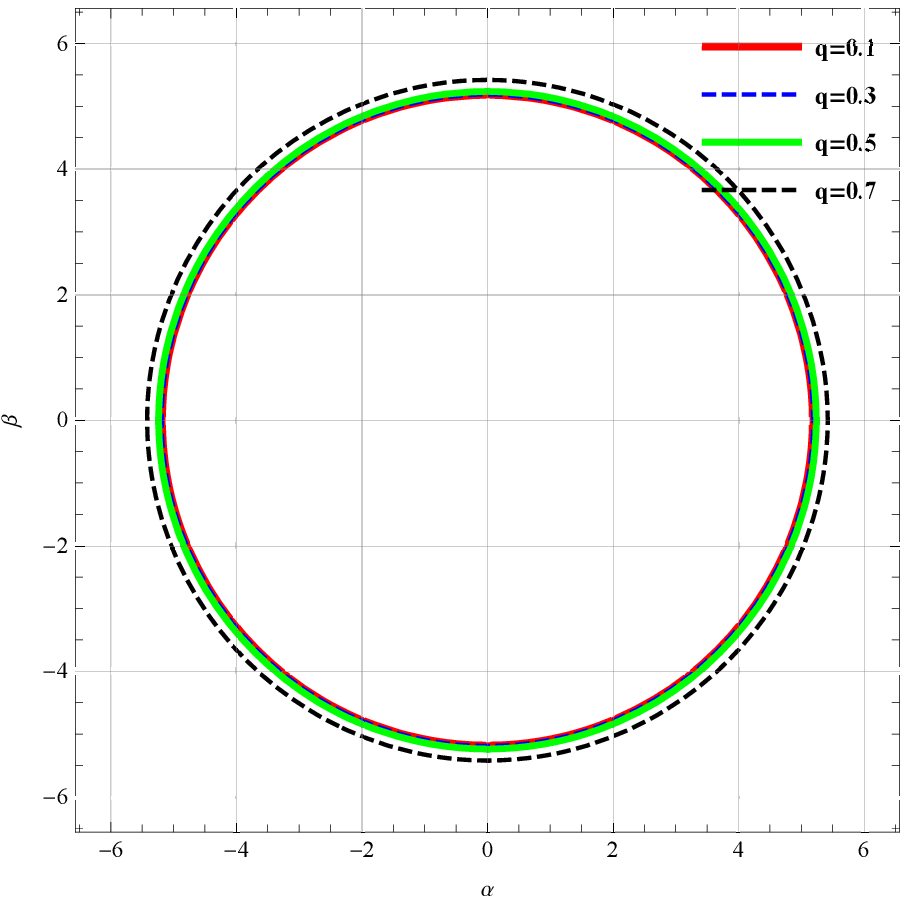}\label{5b}} 
\end{array}$
\end{center}
\caption{In (a), black hole shadow in  $\alpha\beta$-plane for different values of $q$ with $k=0.2$. In (b), black hole shadow in the  $\alpha\beta$-plane  for different values of $q$ with $k=0.4$. Here, $\Lambda = 0.0098$ and $M = 1$.}
\label{fig5}
\end{figure}

The figure \ref{fig5} depicts the dependence of the black hole shadow on charge $q$ and plasma parameter $k$ with cosmological constant $\Lambda = 0.0098$ in celestial plane  for plasma medium. We see that shadow radius increases for an increase in plasma parameter $k$ with changing photon sphere radius. It is also evident from the plots that the size of the black hole shadow rises for an increase in the value of the charge parameter $q$.
  
Next, we shall discuss the effects of  $q$ and $\Lambda$ on the shadow radius  for two different values of $k$.
\section{Effect  of   parameters on shadow radius  in plasma medium}\label{sec8}
In this section, we  study the effects of various parameter on shadow radius  for the plasma media. The shadow radius $R_{s}$ in  the plasma medium is given by
\begin{equation}\label{075}
R_{s}  =\sqrt{ \frac{n^2r^2_{p}}{f(r_{p})+\Lambda n^2r^2_{p}}}.
\end{equation}
It is evident that shadow radius $R_{s}$ depends on the parameters such as   $\Lambda$,   $q$ and $k$.

\begin{figure}[h!]
\begin{center} 
$\begin{array}{cccc}
\subfigure[]{\includegraphics[width=0.5\linewidth]{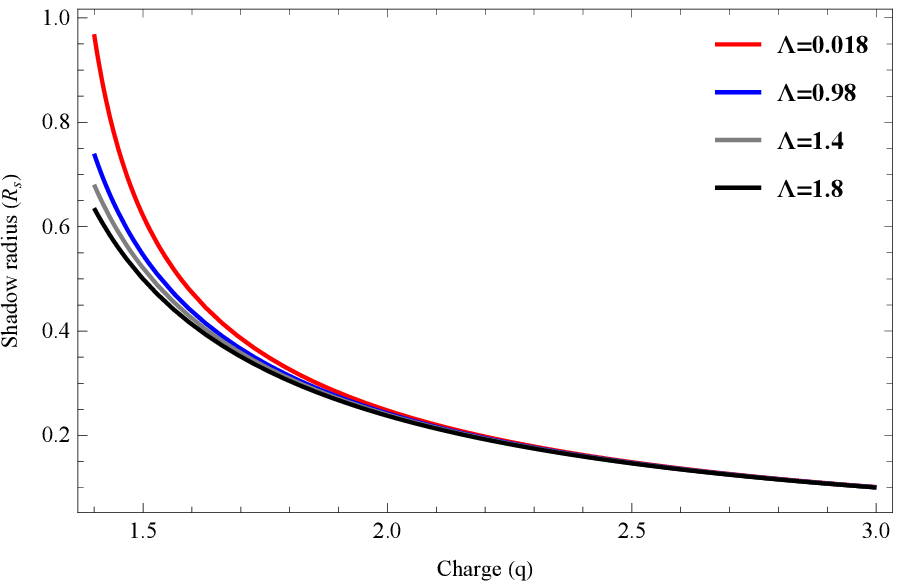}
\label{6a}}
\subfigure[]{\includegraphics[width=0.5\linewidth]{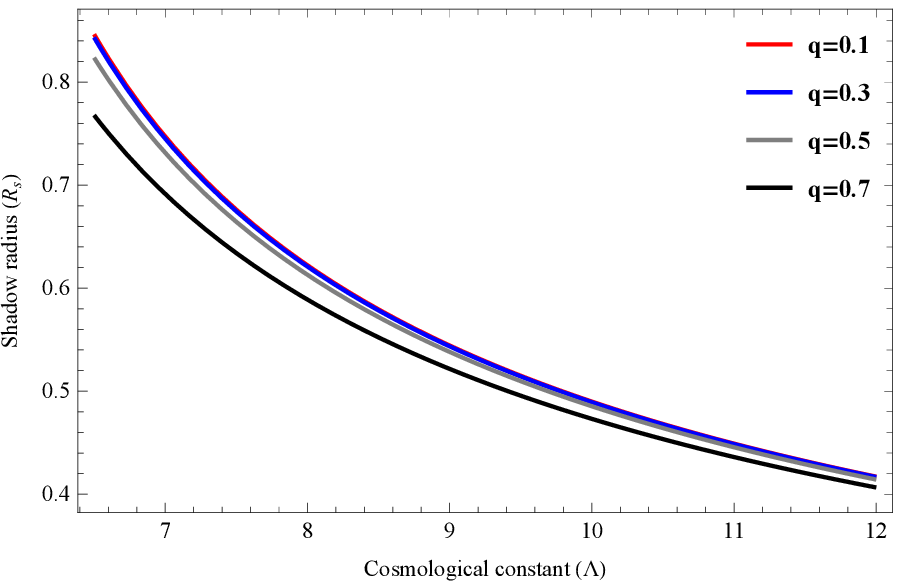}\label{6b}} 
\end{array}$
\end{center}
\caption{In (a),   $R_{s}$ vs. $q$ for changing cosmological constant $\Lambda$. In (b),  $R_{s}$ vs.   $\Lambda$ for changing charge $q$. Here, $k=0.2$ and $M = 1$}
\label{fig6}
\end{figure}

The figure \ref{fig6} depicts how shadow radius for plasma medium depends on parameter like charge $q$ and cosmological constant $\Lambda$ for $k = 0.2$. Here, in \ref{6a}, we see that shadow radius is a decreasing function of charge for fixed values of photon sphere radius. For larger $\Lambda$ values, shadow radius curve decreases sharply. However, in plot \ref{6b}, shadow radius is a decreasing function of $\Lambda$ for fixed photon sphere radius. Moreover, the shadow radius curve decreases for increasing charge parameter $q$. 

\begin{figure}[h!]
\begin{center} 
$\begin{array}{cccc}
\subfigure[]{\includegraphics[width=0.5\linewidth]{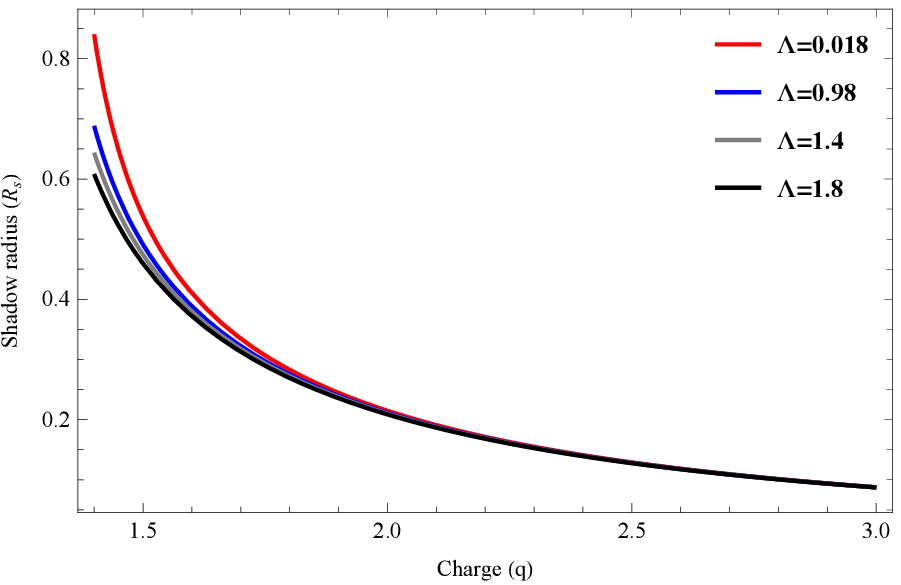}
\label{7a}}
\subfigure[]{\includegraphics[width=0.5\linewidth]{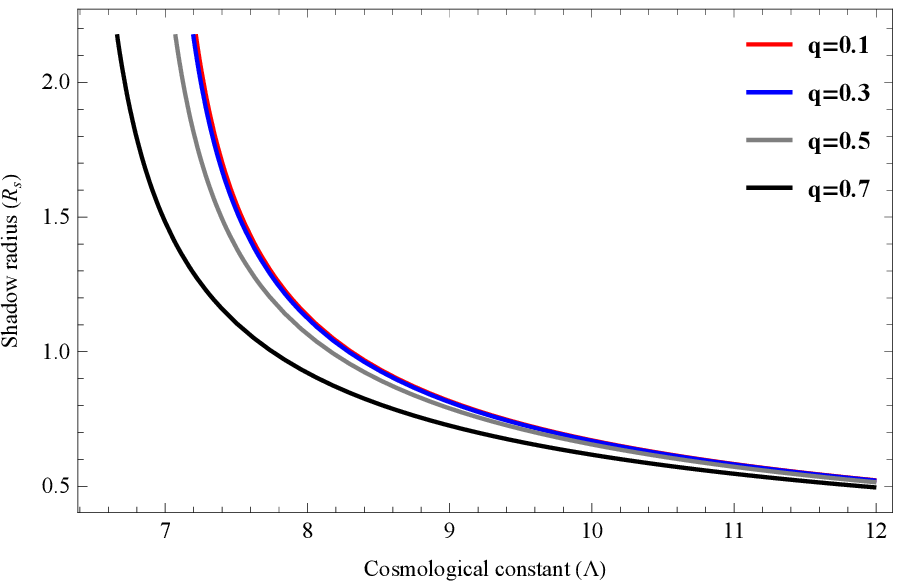}\label{7b}} 
\end{array}$
\end{center}
\caption{In (a),  $R_{s}$ vs. $q$ for changing cosmological constant $\Lambda$. In (b),  $R_{s}$ vs. $\Lambda$ for changing charge $q$. Here, $k=0.4$ and $M = 1$}
\label{fig7}
\end{figure}

The figure \ref{fig7} depicts how shadow radius for plasma medium depends on parameter like charge $q$ and cosmological constant $\Lambda$ with $k = 0.4$. In \ref{7a}, shadow radius is a decreasing function of $q$ for fixed values of photon sphere radius. We also see that shadow radius curve decreases sharply for larger $\Lambda$ values. However, in plot \ref{7b}, shadow radius is a decreasing function of $\Lambda$ for fixed photon sphere radius. Moreover, the shadow radius curve falls for increasing charge parameter $q$. 
  
Both figures \ref{fig6} and \ref{fig7} suggest  that, for a fixed photon sphere radius, shadow radius  is a decreasing function of plasma parameter.

\section{Energy emission rate}\label{sec9}
Here, in the present section, we discuss the energy emission rate of $5D$ Reissner-Nordstr\"om AdS black hole. The
  energy emission rate can be expressed as \cite{s8, s9, s10, s11}
\begin{equation}\label{76}
\frac{d^2}{d\omega dt}Z(\omega)=\frac{2\pi^2\sigma_{lim}}{\exp\left(\frac{\omega}{T_{H}}\right)-1}\omega^3,
\end{equation}
where $Z(\omega)$, $T_{H}$ and $\omega$ are the energy, Hawking temperature and frequency, respectively, for the black hole. Here,  limiting constant   $\sigma_{lim}$ for a spherically symmetric black corresponds to the geometrical cross-section of its photon sphere. The Hawking temperature can be calculated as \cite{s12a}
\begin{equation}\label{77}
T_{H}=\left.\frac{f^{\prime}(r)}{4\pi}\right|_{r=r_{+}}=\frac{2M^2}{\pi r^3_{+}}-\frac{q^4}{\pi r^5_{+}}-\frac{\Lambda r_{+}}{12\pi},
\end{equation}
where $r_{+}$ is the event horizon radius of RN-AdS$_{5}$ black hole.
 
For $5D$ black hole $\sigma_{lim}$ reads \cite{s12, s13, s14}
\begin{equation}\label{78}
\sigma_{lim}=\frac{4\pi R^3_{s}}{3}.
\end{equation}
With this $\sigma_{lim}$, the energy emission rate   becomes
\begin{equation}\label{79}
\frac{d^2}{d\omega dt}Z(\omega)=\frac{8\pi^3R^3_{s}}{3\left(\exp\left(\frac{\omega}{T_{H}}\right)-1\right)}\omega^3.
\end{equation}
In order to study the behavior of the energy emission rate, we plot graphs.
\begin{figure}[h!]
\begin{center} 
$\begin{array}{cccc}
\subfigure[]{\includegraphics[width=0.5\linewidth]{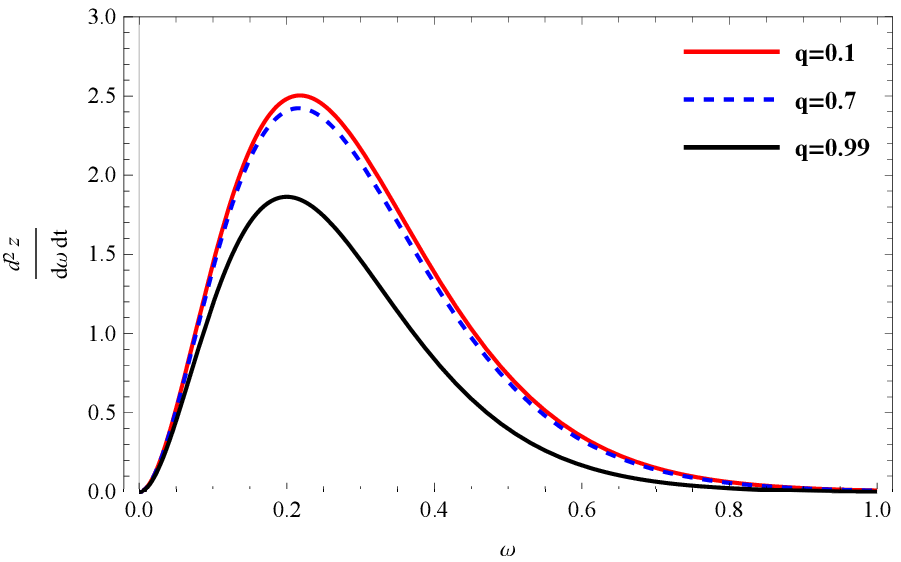}
\label{8a}}
\subfigure[]{\includegraphics[width=0.5\linewidth]{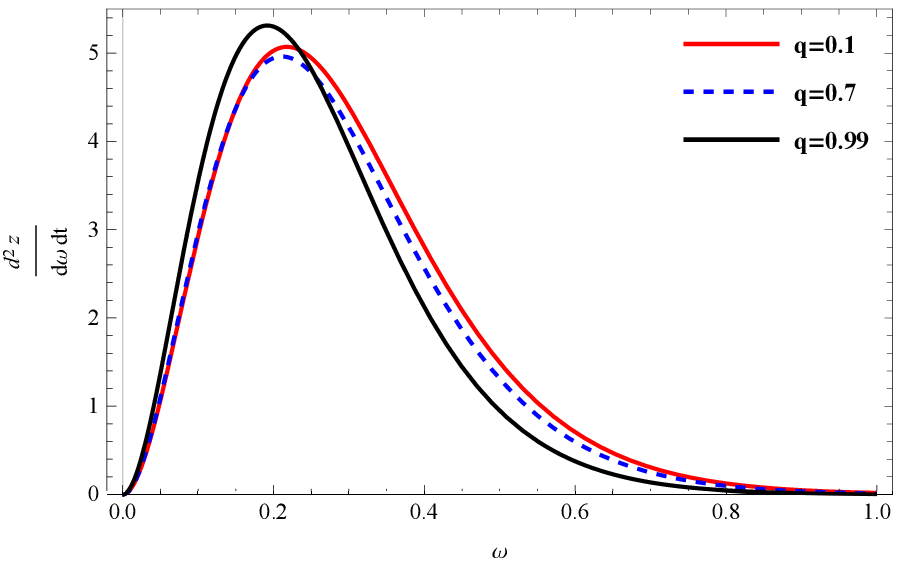}\label{8b}}\\ 
\subfigure[]{\includegraphics[width=0.5\linewidth]{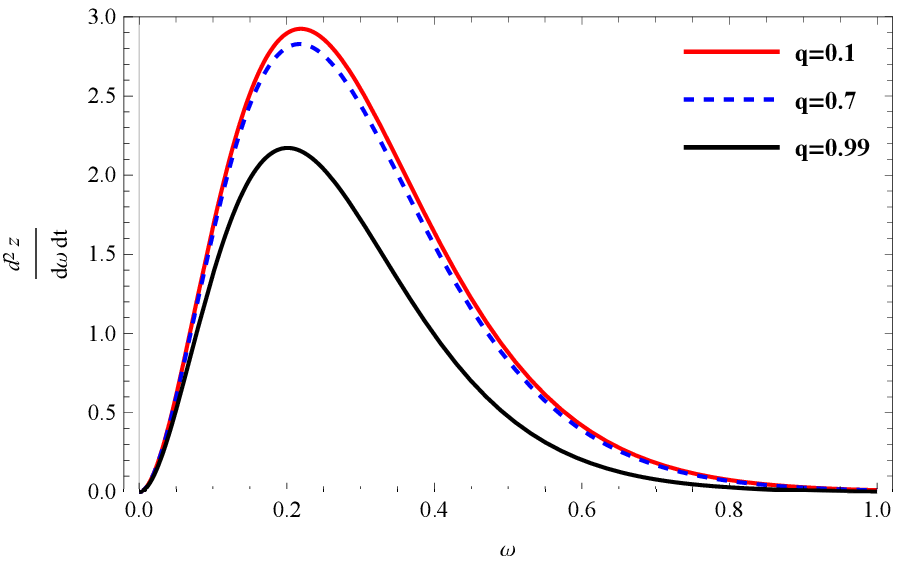}\label{8c}}
\subfigure[]{\includegraphics[width=0.5\linewidth]{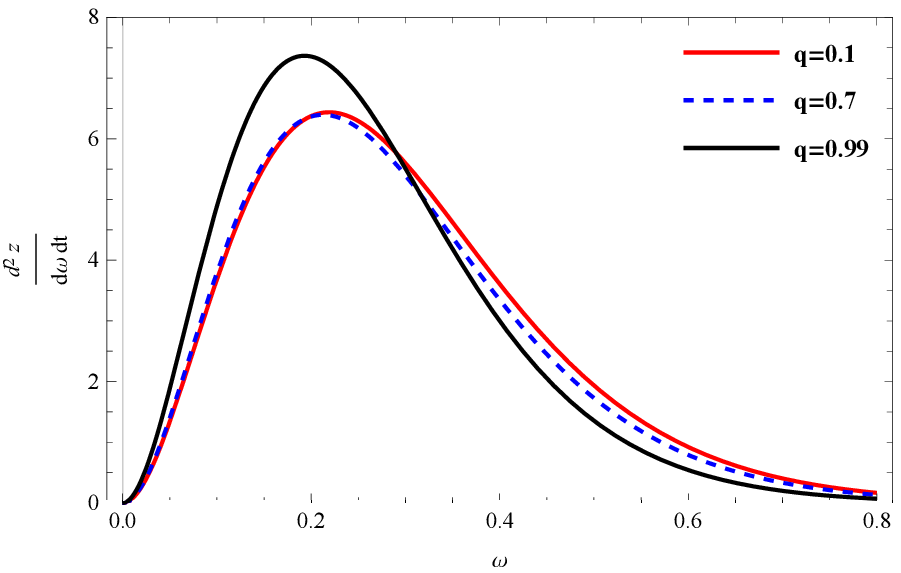}\label{8d}}
\end{array}$
\end{center}
\caption{The energy emission rate vs. frequency   for changing charge $q$ and changing photon sphere radius. In (a) $k = 0$, $\Lambda = 0.018$; in (b) $k = 0.2$, $\Lambda = 0.018$; in (c) $k = 0$, $\Lambda = 0.0098$ and in (d) $k = 0.2$, $\Lambda = 0.0098$. For all cases $M = 1$.}
\label{fig8}
\end{figure}
\begin{figure}[h!]
\begin{center} 
$\begin{array}{cccc}
\subfigure[]{\includegraphics[width=0.5\linewidth]{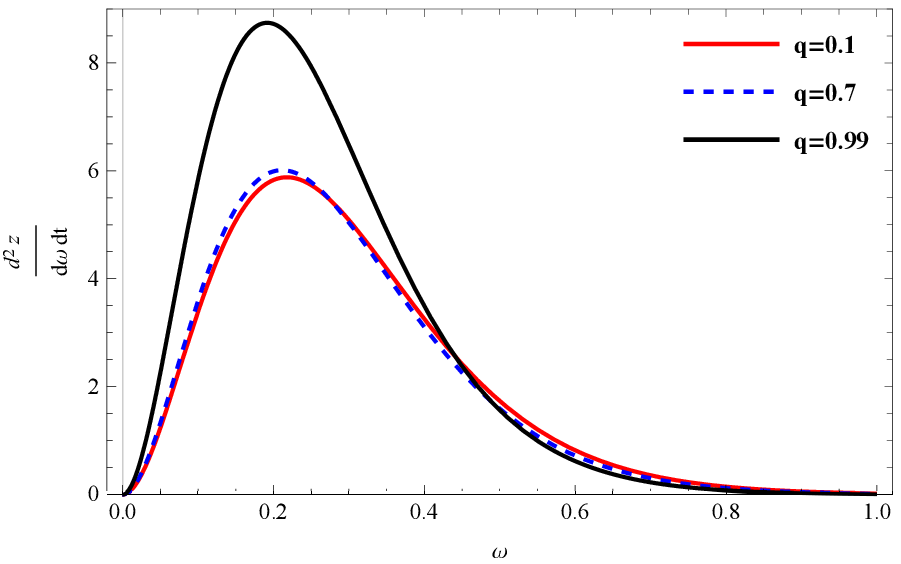}
\label{9a}}
\subfigure[]{\includegraphics[width=0.5\linewidth]{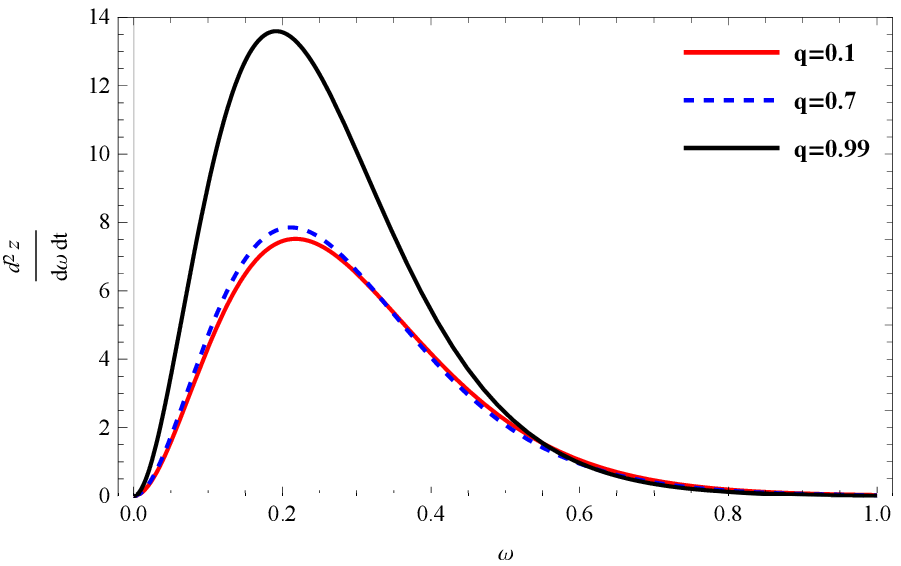}\label{9b}}
\end{array}$
\end{center}
\caption{The energy emission rate vs. frequency  for changing charge $q$ and changing photon sphere radius. In (a) $k = 0.4$, $\Lambda = 0.018$ and in (b) $k = 0.4$, $\Lambda = 0.0098$. For all cases $M = 1$.}
\label{fig9}
\end{figure}
The figure \ref{fig8} depicts the  behavior of energy emission rate $\frac{d^2Z(\omega)}{d\omega dt}$ with respect to frequency $\omega$ for varying charge $q$, varying photon sphere radius with fixed plasma parameter $k = 0, 0.2$. In contrast to non-plasma medium, the energy emission rate increases for increases in charge for plasma medium. The presence of  plasma parameter $k$ increases the energy emission rate. However, the  behavior of energy emission rate   with respect to frequency $\omega$ for varying charge $q$, varying photon sphere radius with fixed plasma parameter $k = 0.4$ is shown in figure \ref{fig9}. Here, it is more clear that the energy emission rate increases for increase  in the values of charge.
\section{Conclusion}\label{sec10}
We now summarize our results here, in the current section. We have investigated the shadow of RN AdS$_{5}$ black
holes for an infinitely distant observer. We have first evaluated the null geodesic equations using the Hamilton-Jacobi equation. Using the boundary condition of unstable circular orbit, we have determined the radius of the photon sphere and then  by using the null geodesics we have obtained the coordinates ($\alpha$, $\beta$) which eventually results the radius of the shadow.  We have tabulated of the various numerical values of photon radius   and shadow radius  for different  charge. We have also done a graphical analysis for the black hole shadow in the  $\alpha\beta$-plane  for different   charges. The plots declared that the radius of the black hole shadow decreases for an increase in the values of both the charge parameter   and  cosmological constant. We have also found that shadow radius depends on the  mass of the black hole also. To check the dependency of the shadow radius on these parameters, we have made a graphical analysis and observed that shadow radius is a decreasing function of charge for fixed photon sphere radius.

 We then introduced a plasma medium in order to investigate their effect    on the unstable circular orbits of null-like particles (photon). It is observed that shadow radius enlarges with the plasma parameter $k$ with varying photon sphere radius. The resulting shadow radius in the plasma medium has dependency on plasma parameter. We have done the graphical analyses for the plasma medium also, which reflect  the effect of various parameters on shadow radius. Finally, we studied the energy emission rate of the  $5D$ RN AdS  black hole and have provided a graphical analysis of energy emission rate with respect to frequency. For future perspective, by utilizing the mathematical optics,   we shall carry forward the present study and  will attempt to distinguish the differences between the black holes acquired from various theories. We also would like to study the consequences of spin parameter on the shadows of RN AdS$_5$ black hole. 

\section*{Acknowledgment}
This research was funded by the Science Committee of the Ministry of Education and Science of the Republic of Kazakhstan (Grant No. AP08052034). 

\section*{Data availability}
 Data sharing is not applicable to this manuscript, as no data sets were generated or analyzed during the current study.

\end{document}